\documentclass[12pt,preprint]{aastex}
\usepackage{graphics,graphicx,rotating,amsmath}

\newcommand{\dt}{{\Delta\hspace{-0.3mm}\theta}}
\newcommand{\db}{{\Delta\hspace{-0.3mm}\beta}}
\newcommand{\dtb}{{\Delta\hspace{-0.3mm}\tilde\beta}}
\newcommand{\Real}[1]{{\rm Re}\left[ #1 \right]}
\newcommand{\tp}{\hspace{-1mm}+\hspace{-1mm}}
\newcommand{\tm}{\hspace{-1mm}-\hspace{-1mm}}

\newcommand{\cH}{{\cal H}}
\newcommand{\bdelta}{{\boldsymbol \delta}}
\newcommand{\bq}{{\mbox{\boldmath$q$}}}

\newcommand{\lr}[1]{\left( #1 \right)}
\newcommand{\slr}[1]{\left[ #1 \right]}

\begin{document}
\title{Elliptical Weighted HOLICs for Weak Lensing Shear Measurement\\
part1:Definitions and isotropic PSF correction}
\author{Yuki Okura\altaffilmark{1}} 
\email{yuki.okura@nao.ac.jp}

\author{Toshifumi Futamase\altaffilmark{2}}
\email{tof@astr.tohoku.ac.jp}

\altaffiltext{1}
 {National Astronomical Observatory of Japan, Tokyo 181-8588, Japan}
\altaffiltext{2}
 {Astronomical Institute, Tohoku University, Sendai 980-8578, Japan}

\begin{abstract}
We develop a new method to estimate gravitational shear  
by adopting an elliptical weight function to measure background galaxy images. 
In doing so, we introduce a new concept of "zero plane" which is an imaginal source plane where shapes of all sources are perfect circles, and regard the intrinsic shear as the result of an imaginal lensing distortion. 
This makes the relation between the observed shear, the intrinsic shear and lensing distortion 
more simple and thus higher-order calculation more easy. 
The elliptical weight function allows us to measure the mutiplemoment of shape of background galaxies more precisely by weighting highly to brighter parts of image and moreover to reduce systematic error due to insufficient expansion of the weight function in the original approach of KSB. 
Point Spread Function(PSF) correction in E-HOLICs methods becomes more complicated than those in KSB methods. In this paper we studied isotropic PSF correction in detail. By adopting the lensing distortion as the ellipticity of the weight function, we are able to show that 
the shear estimation in E-HOLICs method reduces to solve a polynomial in the absolute magnitude of the distortion.
We compare the systematic errors between our approach and KSB using STEP2 simulation. 
It is confirmed that KSB method overestimate the input shear for images with large ellipticities, 
and E-HOLICs correctly estimate the input shear even for such images. 
Anisotropic PSF correction and analysis of real data will be presented in forthcoming paper.
\end{abstract}
\section{Introduction}
Weak gravitation lensing analysis has been widely recognized as a very useful and unique method to study not only mass distribution of the universe, 
but also the cosmological parameters(see for example Mellier 1999, Refregier 2003, van Waerbeke 2003, Munshi 2008). 
However the signal of weak lensing is small and thus accurate method of shear estimation must be required. 
Many method have been proposed so far and the most popular one is KSB method(Kaiser, Squires \& Broadhurst 1995, Luppino \& Kaiser 1997, Hoekstra et al 1998) which measures moments of the galactic light distribution and calculates change of moments by lensing distortion.
The KSB method is successfully applied to many objects(Kaiser \& Squires 1993, Broadhurst et al 2005, Umetsu \& Broadhurst 2007, Okabe \& Umetsu 2008 and Okabe 2010) 
including cosmic shear which is distortions by large scale structure(Bacon et al 2000, Kaiser et al 2000, van Waerbeke et al 2001 and Hamana 2003). 

In particular the cosmic shear attracts much attention recently because of its potentiality to determine the so-called the cosmic equation of state, 
namely the relation between the energy density and pressure of the dark energy which will decide the destination of the universe. 
However the lensing signal of the cosmic shear is much smaller than the signal by cluster weak lensing and thus highly accurate measurement of shear is necessary in order to determine the cosmic equation of state in a percent level. 
There are several sources for noise in the estimation process of lensing distortion which we have to reduce as possible as we can. 
For example, sources have intrinsic ellipticities before affected by lensing distortion. Because we can not divide the estimated distortion into intrinsic ellipticity and lensing distortion, the intrinsic ellipticity behaves as a noise. 
We can reduce this noise by averaging over enough number of the estimated distortions because the intrinsic ellipticities are expected to have random orientations. 
In this way the noise from intrinsic ellipticity is controllable. However there are other noises which are uncontrollable such as noises coming from random photon count, smearing by atmosphere. 
We do not have methods to correct these effects completely, and insufficiency of correction for these effects becomes a systematic noise.
Therefore we need to develop a method which can measure more accurately shape of background galaxies and correct more carefully the effect of the smearing.
The precision and problems of KSB method have been studied (for example, Kuijken 1999, Erben et al. 2001, Hoekstra 2004, Hamana \& Miyazaki 2008).
Then new analysis methods (for example shaplet(Refregier 2003, Refregier \& Bacon 2003, Kuijken et al. 2006) and lensfit(Miller et al 2007 \& Kiching et al 2008))
and improvements of them(Recently, Viola et al. 2010, Melchior et al. 2010) have been developed.

In order to test the accuracy of various shear measurement schemes, some testing program such STEP1 (Heymans et al 2005) and STEP2 (Massey et al 2006) have been developed. For example, the results of STEP2 showed that many of them reach the level of a percent accuracy.   
However, there are two major issues in this result. 
One is that nearly all methods need additional weights and extra corrections to achieve such an accuracy.    
Another is that most of the previous work are interested only in the averaged precisions, namely they are studied the averaged distortion of all images. 
Thus there is a possibility that high accuracy is the result of cancelation between the over-estimate and under-estimate of the distortion for different type of the images. 
Even if we have a set of lens images which has correct average value, 
variance increases(decreases) by overestimate(underestimate) of the shear. 
Especially a correlation between redshift and these over and/or under estimation affects seriously the cosmic shear estimation and the prediction for the cosmic equation of state.

In this paper we develop a new method of the evaluating shear which is based on KSB method, but improves the above two issues. 
Our method is a natural development of our previous studies. 
One is oct-HOLICs (Okura et al 2007 and Okura et al 2008) which estimates higher order distortion "flexions" by using oct-pole components of HOLICs, and another is S2-HOLICs (Okura et al 2009) for precise measurement of shear by using high order spin-2 components and increasing information of image. 
We will generalize the HOLICs approach by adopting an elliptical weight, which we call elliptical weighted HOLICs(E-HOLICs) method. 
We will show that HOLICs method with an elliptical weight is able to measures the lensing distortion 
more accurately by weighting highly to brighter region of image than in the standard KSB method,  
and thus it can reduce effects of systematic error and random noise more effectively than KSB method.

The organization of the paper is as follows. First we give a very brief review of weak lensing in section 2. Then in section 3, 
we introduce a new notations and definitions for HOLICs which make calculation a little easy, 
and then we introduce the concept of "zero plane" which is an imaginal source plane where all sources are perfect circle. 
The source plane is regarded as the result of an imaginal lens mapping from the zero plane.  
This allow us to separate the shear estimation into a process of evaluating ellipticity before smeared by PSF and a process of reducing noise of intrinsic ellipticities by averaging. 
KSB method is examined from our point of view in section 4.
Then E-HOLICs method is presented in section 5.
Tests of KSB method and E-HOLICs method using STEP2 simulation data are shown in section 6.
Finally we summarize our result and give some discussion in section 7.
\section{Weak Lensing convergence, shear and distortion}
A gravitational deflection of light rays can be described by the
lens equation as
\begin{equation}
\label{eq:lenseq}
\beta = \theta - \partial \psi(\theta),
\end{equation}
where $\theta=\theta_1+i\theta_2$ and $\beta=\beta_1+i\beta_2$ are angular positions of the image and source, respectively, 
$\partial = \partial_1 +i\partial_2$ is a complex gradient operator that transforms as a vector, so 
$\partial'=\partial e^{i\phi}$ with $\phi$ being the angle of rotation,
and $\psi(\theta)$ is the effective lensing potential, which
is defined by the two-dimensional Poisson equation as
$\nabla^2\psi(\theta)=2 \kappa(\theta)$, with the lensing convergence.
Here the convergence $\kappa=\Sigma_m \Sigma_{\rm crit}^{-1}$ is the
dimensionless surface mass density projected on the sky, normalized with
respect to the critical surface mass density of gravitational lensing
$\Sigma_{\rm crit} = (c^2D_s)/(4\pi GD_d D_{ds})$,
where $D_d$, $D_s$, and $D_{ds}$ are the angular diameter distances
from the observer to the deflector, from the observer to the source,
and from the deflector to the source, respectively.

The lensing convergence $\kappa$ is expressed as
\begin{equation}
\label{eq:kappa}
\kappa = \frac{1}{2}\partial\partial^* \psi,
\end{equation}
where $^*$ denotes a complex conjugate.
Similarly, the complex gravitational shear is defined as
\begin{equation}
\gamma \equiv \gamma_1+i\gamma_2= \frac{1}{2}\partial\partial \psi
\end{equation} 
which has a property of spin-2.
Note that a quantity is said to have spin-s if it has the same value 
after rotation by $2\pi/s$.  

Because these convergence and shear have a relation in Fourier space as
\begin{equation}
\hat \kappa({\mbox{\boldmath$k$}})=\frac{k_1^2-k_2^2-2ik_1k_2}{k_1^2+k_2^2}\hat \gamma({\mbox{\boldmath$k$}}),
\end{equation} 
therefore mass distribution of lens object can be obtained from shear distribution.

Here we define a reduced shear $g$ and a complex distortion $\bdelta$ which appear in following sections as 
\begin{eqnarray}
g&\equiv&\frac{\gamma}{1-\kappa}\\
\bdelta&\equiv&\frac{2g}{1+|g|^2}.
\end{eqnarray} 

\section{New Notations, New Definitions and Zero plane}
In this section, first we introduce complex forms in displacements from centroid of image and moments of image.
These complex form makes equations much simpler and higher-order calculations easier. 
Next, we introduce "Zero image" which is imaginary circular image and becomes "source" by an imaginary distortion. 
In estimating shear, intrinsic ellipticities of background galaxies complicate equations, because an ellipticity of distorted image is not a simple sum of the intrinsic ellipticity and distortion.
By treating the intrinsic ellipticity as an imaginary distortion from the zero image, the relation between the observed shear, the intrinsic shear and the lensing distortion  becomes a simple form(see eq.(31) below), 
and we are able to divide the shear estimation process into the estimation of individual distortions(which contains intrinsic ellipticity) and the reduction of noise of intrinsic ellipticities by averaging.

\subsection{Complex Notation}
We use complex notation for displacements and moments of image which are expressed by subscripts in KSB method.
\subsubsection{Complex Displacement}
Let $\bar\theta\equiv\bar\theta_1+i\bar\theta_2$ be a centroid of image and a displacement from this point is notated as 
\begin{eqnarray}
\dt\equiv\theta-\bar\theta\equiv(\theta_1-\bar\theta_1)+i(\theta_2-\bar\theta_2).
\end{eqnarray}
Then a combination of displacements is written as
\begin{eqnarray}
\dt^N_M\equiv\dt^{\frac{N+M}{2}}{\dt^*}^{\frac{N-M}{2}},
\end{eqnarray}
where "N" means order of length and "M" means a spin number.

\subsubsection{Complex Moments and HOLICs}
Complex moments of an image without weight function is defined as follows. 
\begin{eqnarray}
Z^N_M(I)\equiv\int d^2\theta I(\theta) \dt^N_M,
\end{eqnarray}
where  $I(\theta)$ is the brightness distribution. Similarly the complex moments with weight function $W$ which has an ellipticity value same as $\bdelta$ is 
defined as
\begin{eqnarray}
Z^N_M(I,\bdelta)\equiv\int d^2\theta I(\theta) \dt^N_M W\lr{\frac{\dt^2_0-\Real{\bdelta^*\dt^2_2}}{\sigma^2}},
\end{eqnarray}
where "N" and "M" mean order of length and spin number respectively.
Because distortion makes ellipticity, distortion and ellipticity have same meaning in E-HOLICs method each other.
For example, comparing moments with weight function with KSB notations becomes
\begin{eqnarray}
Z^2_2(I,0)&=&Q_{11}-Q_{22}+2iQ_{12}\\
Z^2_0(I,0)&=&Q_{11}+Q_{22},
\end{eqnarray}
where $Q_{ij}$ is the components of quadrupole moment in KSB notation.

In addition, we introduce HOLICs s$\cH^N_M$ which is the complex moment $Z^N_M(I)$ normalized by $Z^L_K(I)$ as follow and notate as
\begin{equation}
\cH^N_M(I,Z^L_K)\equiv\frac{Z^N_M(I)}{Z^L_K(I)}
\end{equation}
Similarly we introduce elliptically weighted HOLICs(E-HOLICs) as
\begin{equation}
\cH^N_M(I,Z^L_K,\bdelta)\equiv\frac{Z^N_M(I,\bdelta)}{Z^L_K(I,\bdelta)}.
\end{equation}
For notational simplicity, we write a combination of HOLICs as 
\begin{eqnarray}
\frac{\cH^N_M(I,Z^L_K,\bdelta)+\cH^P_O(I,Z^L_K,\bdelta)}{\cH^R_Q(I,Z^L_K,\bdelta)}\equiv\slr{\frac{\cH^N_M+\cH^P_O}{\cH^R_Q}}(I,Z^L_K,\bdelta)
\end{eqnarray}


\subsection{Source, Image and Zero image without weights}
Here, we define an imaginal brightness distribution "Zero image" which allows us to calculate distortion easily, and present the relations between "Zero image", "Source" and "Image".
\subsubsection{Source and Image}
First, we present a relation between Source and Image which can be written by convergence and shear.

A relation between $\db$ and $\dt$ which are complex displacements in source plane and image plane,  respectively, can be written in matrix expression as follows,
\begin{eqnarray}
\left( \begin{array}{c} \db_1 \\ \db_2 \end{array}\right)
=\left( \begin{array}{cc} 1 \tm {\kappa} \tm {\gamma^L_1} &  \tm {\gamma^L_2} \\
						  \tm {\gamma^L_2}  & 1 \tm {\kappa} \tp {\gamma^L_1} \end{array}\right)
\left( \begin{array}{c} \dt_1 \\ \dt_2 \end{array}\right)
=(1 \tm \kappa)\left( \begin{array}{cc} 1 \tm g^L_1 &  \tm g^L_2 \\
						  \tm g^L_2  & 1 \tp g^L_1 \end{array}\right)
\left( \begin{array}{c} \dt_1 \\ \dt_2 \end{array}\right),
\end{eqnarray}
where $g^L=\gamma^L/(1-\kappa)$ is reduced shear, 
and this relation can be written in our complex notation easily as
\begin{eqnarray}
\db=(1-\kappa)\dt-\gamma^L\dt^*=(1-\kappa)\lr{\dt-g^L\dt^*}.
\end{eqnarray}

By writing brightness distributions of Source and Image as $I^S(\beta)$ and $I^L(\theta)$, respectively,  we obtain the following relation for complex moments of source and image. 
\begin{eqnarray}
Z^N_M(I^S)\equiv\int d^2\beta I^S(\beta)\db^N_M = (1-\kappa)^{N+2}(1-|g^L|^2)\int d^2\theta I^L(\theta)(\dt-g^L\dt^*)^N_M.
\end{eqnarray}

\subsubsection{Source and Zero image}
Next, we present a relation between Source and imaginal Zero image ,
where $I^Z$ and $\dtb$ are a brightness distribution and a complex displacement of a zero image, respectively.
The zero image is defined as a brightness distribution which has no non-spin-0 moments,
and we can relate zero plane and source plane by imaginal shear $\gamma^I$ (= $g^I$, there are no imaginal convergence) as
\begin{eqnarray}
\dtb=\db-\gamma^I\db^*.
\end{eqnarray}
Here we assumed that there is no imaginal convergence, so the imaginal reduced shear is defined as 
\begin{eqnarray}
g^I\equiv\gamma^I.
\end{eqnarray}

From the above definitions, the complex moment $Z^2_2(I^Z)$ of a zero image can be calculated as
\begin{eqnarray}
\label{eq:LensCS}
0=Z^2_2(I^Z)&\equiv&\int d^2\tilde\beta \,  I^Z(\tilde\beta)\dtb^2_2 \\
&=&\lr{1-|g^I|^2}\int d^2\beta I^S(\beta)\lr{\db^2_2-2g^I\db^2_0+g^{I2}{\db^2_2}^*}\\
&=&\lr{1-|g^I|^2}\lr{Z^2_2(I^S) \tm 2g^IZ^2_0(I^S) \tp |g^I|^2Z^2_2(I^S)}\\
&=&\lr{1-|g^I|^2}\lr{1+|g^I|^2}\lr{Z^2_2(I^S)-\frac{2g^I}{1+|g^I|^2}Z^2_0(I^S)}\\
&\equiv&\lr{1-|g^I|^4}\lr{Z^2_2(I^S)-\bdelta^IZ^2_0(I^S)}.
\end{eqnarray}
where we have used the fact that $g^{I2}{Z^{2*}_2(I^S)}=|g^I|^2Z^2_2(I^S)$, namely $Z^2_2(I^S)$ and  $g^I$ have same phase angle because $Z^2_2(I^S)$ is distorted by $g^I$ from a zero image.
Then we can obtain ellipticity easily as,
\begin{eqnarray}
\cH^2_2(I^S,Z^2_0)=\bdelta^I.
\end{eqnarray}
Therefore we can simply use $\delta^I$ and $g^I$ instead of intrinsic ellipticity $\cH^2_2(I^S,Z^2_0)$.

\subsubsection{Image and Zero image, and Lensing shear}
From the above relations, Image can be obtained by two distortions by $g^I$ and $g^L$ from Zero image.
Complex displacement in image plane is distorted as 
\begin{eqnarray}
\dtb&=&(\db-g^I\db^*)=(1-\kappa)\lr{\dt-g^L\dt^*-g^I(\dt^*-g^{L*}\dt)}\\
&=&(1-\kappa)\lr{1-g^{L*}g^I}\lr{\dt-g^C\dt^*},
\end{eqnarray}
where 
\begin{eqnarray}
g^C\equiv\frac{g^I+g^L}{1-g^Ig^{L*}}
\end{eqnarray}
is the combined shear.
From similar calculation of eq.(\ref{eq:LensCS}), we obtain
\begin{eqnarray}
\label{eq:LensCI22}
\cH^2_2(I^L,Z^2_0)=\bdelta^C\equiv\frac{2g^C}{1+|g^C|^2}.
\end{eqnarray}
Because $g^I$ has random phase angle,
Lensing shear $g^L$ can be obtained as shear which satisfies following equation
\begin{eqnarray}
\label{eq:gICL}
0 = \left< g^I \right> = \left< \frac{g^C - g^L}{1 + g^{L*}g^C} \right>.
\end{eqnarray}
This is our basic relation. 

By introducing the Zero plane, we can divide lensing analysis into two steps which are 
the evaluation of  $g^C$ and then the reduction of noise in intrinsic shear by eq.(\ref{eq:gICL}).
Because we don't use any assumption and the property of E-HOLICs for introducing eq.(\ref{eq:gICL}),
this equation can be used for any methods of weak lensing shear estimating.
Therefore our purpose is to obtain $g^C$ in this paper.
And here and after, $g=g^C$ and $\bdelta=\bdelta^C$ for simply notation.

\subsubsection{Shear Distortion for Higher order moments}
We demonstrated that the effects of lensing distortion for order-2 spin-2 moment is $\cH^2_2(I^L,Z^2_0)=\bdelta$.
From similar calculation of eq.(\ref{eq:LensCI22}), relations between distortion and higher order HOLICs are obtained as 
\begin{eqnarray}
\label{eq:distHighH}
\cH^2_2(I,Z^2_0)&=&\bdelta\\
\cH^4_2(I,Z^4_0)&=&\frac{3         }{2+ \delta^2}\bdelta  \\
\cH^4_4(I,Z^4_0)&=&\frac{3         }{2+ \delta^2}\bdelta^2\\
\cH^6_2(I,Z^6_0)&=&\frac{4+\delta^2}{2+3\delta^2}\bdelta  \\
\cH^6_4(I,Z^6_0)&=&\frac{5         }{2+3\delta^2}\bdelta^2\\
\cH^6_6(I,Z^6_0)&=&\frac{5         }{2+3\delta^2}\bdelta^3,
\end{eqnarray}
where $\delta=|\bdelta|$.
These equations will be used in forthcoming papers.

\section{KSB method using zero plane in new notation}
In this section we review the standard KSB method using zero plane and in our new notation.
For simplicity we suppose $\bar\theta=0$ (e.g. $\theta^N_M=\dt^N_M$ ).
The detailed explanation and calculation can be seen in Bartelmann and Schneider 2001.
\subsection{Weight function}
\label{sec:weight}
In our approach it is natural to assume a circular weight function in the zero plane rather than in the image plane as in the original KSB method.
This will give us the unaveraged expression for PSF corrected shear (see equation.(\ref{eq:deltaKSB}))  and the lensing shear will be obtained by averaging using equation (\ref{eq:gICL}).
The relation between weighted complex moments and complex distortion in our case becomes as
\begin{eqnarray}
Z^N_M(I^Z,0)
&=&\int d^2\tilde\beta I^Z(\tilde\beta) \tilde\beta^N_M W\lr{\frac{\tilde\beta^2_0}{\tilde\sigma^2}}\\
=(1-&\kappa &)^{N+2}(1-|g|^2)\int d^2\theta I^L(\theta) \lr{\theta-g\theta^*}^N_M W\lr{\frac{\theta^2_0-\Real{\bdelta^*\theta^2_2}}{\sigma^2}}.
\end{eqnarray}
By calculating complex moment $Z^2_2$, a relation between ellipticity and distortion can be obtained as
\begin{eqnarray}
\label{eq:Weightdelta}
0&=&Z^2_2(I^Z,0)=\int d^2\tilde\beta I^Z(\tilde\beta) \tilde\beta^2_2 W\lr{\frac{\tilde\beta^2_0}{\tilde\sigma^2}}\\
&=&(1-\kappa)^4(1-|g|^2)\int d^2\theta I^L(\theta) \lr{\theta-g\theta^*}^2_2 W\lr{\frac{\theta^2_0-\Real{\bdelta^*\theta^2_2}}{\sigma^2}}\\
&=&(1-\kappa)^4(1-|g|^2)\lr{Z^2_2(I^L,\bdelta)-2g Z^2_0(I^L,\bdelta)+g^2Z^{2*}_2(I^L,\bdelta)}\\
&=&(1-\kappa)^4(1-|g|^2)(1+|g|^2)\lr{Z^2_2(I^L,\bdelta)-\bdelta Z^2_0(I^L,\bdelta)}\\
&\approx & (1-\kappa)^4(1-|g|^4)\lr{Z^2_2(I^L,0)-\frac{\bdelta}{2\sigma^2}{Z'}^4_0(I^L,0)-\frac{\bdelta^*}{2\sigma^2}{Z'}^4_4(I^L,0)-\bdelta Z^2_0(I^L,0)},
\end{eqnarray}
where we have expanded the elliptical weight function and kept only 1st order in $\delta$ because KSB method use circular weight function.
\begin{eqnarray}
W\lr{\frac{\theta^2_0-\Real{\bdelta^*\theta^2_2}}{\sigma^2}}\approx W\lr{\frac{\theta^2_0}{\sigma^2}}-\frac{\bdelta^*\theta^2_2}{2\sigma^2}W'\lr{\frac{\theta^2_0}{\sigma^2}}-\frac{\bdelta\theta^{2*}_2}{2\sigma^2}W'\lr{\frac{\theta^2_0}{\sigma^2}},
\end{eqnarray}
and $Z'$ is the complex moment with weight function $\partial W(x)/\partial x$ instead of $W(x)$.
Therefore,
\begin{eqnarray}
\label{eq:WeightdeltaH}
\cH^2_2(I^L,Z^2_0,0)&\approx&\lr{1+\frac{1}{2\sigma^2}{\cH'}^4_0(I^L,Z^2_0,0)}\bdelta+\lr{\frac{1}{2\sigma^2}{\cH'}^4_4(I^L,Z^2_0,0)}\bdelta^*\\
	&\equiv&C^{KSB}_0(I^L)\bdelta+C^{KSB}_4(I^L)\bdelta^*,
\end{eqnarray}
where $\cH'$ is defined as
\begin{eqnarray}
{\cH'}^N_M(I^L,Z^K_L,0)\equiv\frac{{Z'}^N_M(I^L,0)}{Z^K_L(I^L,0)}.
\end{eqnarray}
A complex distortion is thus calculated as follows.
\begin{eqnarray}
\label{eq:H22lens}
\bdelta^{KSB}\approx \frac{\cH^2_2(I^L,Z^2_0,0)C_0(I^L,0)-\cH^{2*}_2(I^L,Z^2_0,0)C_4(I^L,0)}{|C_0(I^L,0)|^2-|C_4(I^L,0)|^2}.
\end{eqnarray}
If we consider only 1st order of combinations in $C$ and $H^N_M$, 
we can neglect this terms like $H^{2*}_2C_4$.
Therefore we can obtain shear simply as
\begin{eqnarray}
\bdelta^{KSB}\approx\frac{\cH^2_2(I^L,Z^2_0,0)}{C_0(I^L,0)}
\end{eqnarray}

Because the approximation in eq.(\ref{eq:H22lens}) is effective under the assumption of small $\bdelta$,
the equation has systematic error in high $\bdelta$.
For example, if image and weight function are Gaussian with same size and have ellipticity $\bdelta_{true}$ and 0, respectively, we can calculate complex moments as
\begin{eqnarray}
Z^N_M=\int d^2\theta \theta^N_M e^{-\theta^2_0/\sigma^2}e^{-\theta^2_0/\sigma^2}
\end{eqnarray}
and obtain
\begin{eqnarray}
\cH^2_2&=&\frac{\bdelta_{true}}{2}\\
C^{KSB}_0&=&\frac{8-5\delta_{true}^2}{4(4-\delta_{true}^2)}\\
C^{KSB}_4&=&\frac{-3\bdelta^2_{true}}{4(4-\delta_{true}^2)}.
\end{eqnarray}
Therefore the estimated distortion $\bdelta_{estimate}$ is derived as
\begin{eqnarray}
\label{eq:syserrorWeight}
\bdelta_{estimate}= \bdelta_{true}\frac{1-\delta_{true}^2/4}{1-\delta_{true}^2}.
\end{eqnarray}
Because KSB method corrects only the numerator of $\cH^2_2$,
one can over estimate the distortion in  general.

\subsection{PSF correction}
Because the observed images are smeared by atmospheric turbulence and so on, we must correct 
these effect to obtain true distortions.
This effect can be expressed by Point Spread Function(PSF) and
 can be divided into isotropic PSF which reduce ellipticity of lensed image and anisotropic PSF which add a extra ellipticity.
KSB method corrects these two effects using moments of star images which is smeared by same PSF.

An observed image $I^{obs}(\theta)$ is smeared by PSF $P(\theta)$ from lensed image  $I^L(\beta)$ and 
expressed as
\begin{eqnarray}
\label{eq:PSF}
I^{obs}(\theta)=\int d\beta I^L(\beta)P(\theta-\beta). 
\end{eqnarray}
KSB assumes that PSF is expressed as a convolution of an isotropic part $P^{iso}$ and an anisotropic part of $q$ as follows.
\begin{eqnarray}
P(\theta)=\int d^2\psi P^{iso}(\psi)q(\theta-\psi).
\end{eqnarray}
The moments of anisotropic PSF is assumed to satisfy the following relations. 
\begin{eqnarray}
\label{eq:qeq}
\bq^0_0&=&1,\\
\bq^1_1&=&0,\\
\bq^N_0&=&0.
\end{eqnarray}
where moments of anisotropic PSF are defined as 
\begin{equation}
\int d^2\theta \theta^N_Mq(\theta)=\bq^N_M,
\end{equation}

Let $I^{iso}$ be an image which is smeared by only isotropic PSF as 
\begin{eqnarray}
\label{iso}
I^{iso}(\theta)=\int d^2\beta I^L(\beta)P^{iso}(\theta-\beta), 
\end{eqnarray}
then eq.(\ref{eq:PSF}) becomes
\begin{eqnarray}
\label{obsiso}
I^{obs}(\theta)=\int d^2\psi I^{iso}(\psi)q(\theta-\psi).
\end{eqnarray}
\subsubsection{Anisotropic PSF correction}
Here, we explain a method of anisotropic PSF correction.

Moments of $I^{obs}$ are described by integral of $I^{iso}$ and $q$ as
\begin{eqnarray}
\int d^2\theta f(\theta)I^{obs}(\theta)&=&\int d^2\theta \int d^2\psi f(\theta)I^{iso}(\psi)q(\theta-\psi)\\
&=& \int d^2\phi \int d^2\psi f(\phi+\psi) I^{iso}(\psi)q(\phi) \\
f(\theta)&=&\theta^N_MW\lr{\frac{\theta^2_0-\Real{\bdelta^*\theta^2_2}}{\sigma^2}}.
\end{eqnarray}
Expanding $f(\phi+\psi)$ up to 1st order of $\delta$,  $Z^2_2(I^{obs},0)$ is obtained as
\begin{eqnarray}
Z^2_2(I^{obs},0) &=& \slr{Z^2_2+\bq^2_2\lr{Z^0_0+\frac{2}{\sigma^2}{Z'}^2_0+\frac{1}{2\sigma^4}{Z''}^4_0}+\bq^{2*}_2{Z''}^4_4}(I^{iso},0),
\end{eqnarray}
Then we have an approximated expression for $\cH^2_2(I^{obs},0)$. 
\begin{eqnarray}
\cH^2_2(I^{obs},&Z^2_0&,0) \equiv \slr{\cH^2_2+\bq^2_2\lr{\cH^0_0+\frac{2}{\sigma^2}{\cH'}^2_0+\frac{1}{2\sigma^4}{\cH''}^4_0}+\bq^{2*}_2{\cH''}^4_4}(I^{iso},Z^2_0,0)\\
	&\approx& \cH^2_2 (I^{iso},Z^2_0, 0)+ \slr{\bq^2_2\lr{\cH^0_0+\frac{2}{\sigma^2}{\cH'}^2_0+\frac{1}{2\sigma^4}{\cH''}^4_0}+\bq^{2*}_2{\cH''}^4_4}(I^{obs},Z^2_0,0)\\
	&\equiv& \cH^2_2(I^{iso},Z^2_0,0)+\bq^2_2P^{KSB}_0(I^{obs})+\bq^{2*}_2P^{KSB}_4(I^{obs}).
\end{eqnarray}
Similar calculation is applied for star image $I^*$ and we obtain 
\begin{eqnarray}
\cH^2_2(I^{*obs},Z^2_0,0) = \bq^2_2P^{KSB}_0(I^{*obs})+\bq^{2*}_2P^{KSB}_4(I^{*obs})
\end{eqnarray}
This will give us the following expression for $\bq^2_2$.
\begin{eqnarray}
\bq^2_2=\frac{\cH^2_2(I^{*obs},Z^2_0,0)P^{KSB}_0(I^{*obs})-\cH^{2*}_2(I^{*obs},Z^2_0,0)P^{KSB}_4(I^{*obs})}{|P^{KSB}_0(I^{*obs})|^2-|P^{KSB}_4(I^{*obs})|^2}.
\end{eqnarray}
If we use 1st order of the above equation, $\cH^{2*}_2P^{KSB}_4$ may be neglected, and we have
\begin{eqnarray}
\bq^2_2=\frac{\cH^2_2(I^{*obs},Z^2_0,0)}{P^{KSB}_0(I^{*obs})}.
\end{eqnarray}
where $P^{KSB}_0$ is tr$P$ which is often used instead of matrix $P$ in weak lensing analysis with KSB method.

\subsubsection{Isotropic PSF correction}
Next, we explain the method of isotropic PSF correction which reduce ellipticity of images.
Let $\tilde I(\tilde\beta)$ be a brightness distribution in zero plane. According to the conservation of brightness, this coincides with $I^{iso}(\theta)$ in image plane, 
and we can describe as
\begin{eqnarray}
I^{iso}(\theta)=\tilde I(\tilde\beta) = \int d^2\tilde\psi I^Z(\tilde\psi)\tilde P(\tilde\beta-\tilde\psi).
\end{eqnarray}
Because $\tilde P$ is not isotropic function, we write $\tilde P$ in the following way. 
\begin{eqnarray}
\tilde P(\tilde \beta)=\int d^2\tilde \psi \tilde P^{iso}(\tilde \psi)\tilde \bq(\tilde \beta-\tilde \psi).
\end{eqnarray}
Let $\tilde I^{iso}$ be a brightness distribution which is smeared by only $ \tilde P^{iso}(\tilde \psi)$,
and $\tilde I(\tilde\beta)$ can be described as
\begin{eqnarray}
\tilde I(\tilde\beta) = \int d^2\tilde\psi \tilde I^{iso}(\tilde\psi)\tilde \bq(\tilde\beta-\tilde\psi).
\end{eqnarray}
Because the anisotropic part $\tilde \bq$ is made by shear, we can assume that $\tilde \bq^0_0$ and $\tilde \bq^2_2$ are dominant in $\tilde \bq$.
Therefore we can correct anisotropic part similarly as above, so we obtain 
\begin{eqnarray}
0=\cH^2_2(\tilde I^{iso},Z^2_0,0) &\approx&\cH^2_2(\tilde I,Z^2_0,0)-P^{KSB}_0(\tilde I)\tilde \bq^2_2-P^{KSB}_4(\tilde I)\tilde \bq^{2*}_2\\
&\approx&\cH^2_2(\tilde I,Z^2_0,0)-P^{KSB}_0(I^{obs})\tilde \bq^2_2-P^{KSB}_4(I^{obs})\tilde \bq^{2*}_2.
\end{eqnarray}
We need the relation between $\cH^2_2(\tilde I,Z^2_0,0)$ and $\cH^2_2(I^{iso},Z^2_0,0)$ 
which is obtained by similar calculation as eq.(\ref{eq:WeightdeltaH}).
\begin{eqnarray}
\cH^2_2(\tilde I,Z^2_0,0)&\approx& \cH^2_2(I^{iso},Z^2_0,0)- C^{KSB}_0(I^{iso})\bdelta- C^{KSB}_4(I^{iso})\bdelta^*\\
&\approx& \cH^2_2(I^{iso},Z^2_0,0)-C^{KSB}_0(I^{obs})\bdelta-C^{KSB}_4(I^{obs})\bdelta^*.
\end{eqnarray}
Thus  $\cH^2_2(I^{iso},Z^2_0,0)$ can be described by $\bdelta$ and $\tilde \bq^2_2$ as follows.
\begin{eqnarray}
\label{eq:shearKSB}
\cH^2_2(I^{iso},Z^2_0,0)\approx C^{KSB}_0(I^{obs})\bdelta +C^{KSB}_4(I^{obs})\bdelta^* +P^{KSB}_0(I^{obs})\tilde \bq^2_2 +P^{KSB}_4(I^{obs})\tilde \bq^{2*}_2.
\end{eqnarray}
$\tilde \bq^2_2$ can be obtained from star images as 
\begin{eqnarray}
0&=&\cH^2_2(I^{*iso},Z^2_0,0)\approx C^{KSB}_0(I^*)\bdelta +C^{KSB}_4(I^*)\bdelta^* +P^{KSB}_0(I^*)\tilde \bq^2_2 +P^{KSB}_4(I^*)\tilde \bq^{2*}_2\\
\tilde \bq^2_2&=&-\slr{\frac{\lr{C^{KSB}_0P^{KSB}_0-C^{KSB*}_4P^{KSB}_4}\bdelta+\lr{C^{KSB}_4P^{KSB}_0-C^{KSB}_0P^{KSB}_4}\bdelta^*}{|P^{KSB}_0|^2-|P^{KSB}_4|^2}}(I^*),
\label{eq:tqKSB}
\end{eqnarray}
Substituting this expression to eq.(\ref{eq:shearKSB}),  we can estimate complex distortion. 
Especially in 1st order, we can use the following simple equation to estimate shear.  
\begin{eqnarray}
\label{eq:deltaKSB}
\bdelta \approx \bdelta^{KSB} \equiv \frac{\cH^2_2(I^{iso},Z^2_0,0)}{C^{KSB}_0(I^{obs})-C^{KSB}_0(I^*)\frac{P^{KSB}_0(I^{obs})}{P^{KSB}_0(I^*)}}.
\end{eqnarray}
\section{Elliptical weighted HOLICs}
In this section, we demonstrate an improvement for the overestimation of shear  
in KSB by adopting an elliptical weight function. 
In this paper we will not consider anisotropic PSF correction which will be treated in the forthcoming paper.
\subsection{Motivation and Merits of using E-HOLICs}
The one of the major concern in weak lensing is to measure the cosmic equation of state by cosmic shear. It is needless to say that this will require us to have highly accurate method to evaluate the gravitational shear. 
Although KSB method is very powerful and convenient,  it uses only the first-order  expansion of the weight function as we demonstrated in the previous sections, and thus does not have enough accuracy in 
such a purpose. For example, KSB has $7.5\%$ bias if the lensed object has ellipticity 0.3 which is typical value of intrinsic ellipticities.
Moreover, the use of a circular weight for elliptical images in KSB method is not an 
efficient and accurate measurement for galaxy shape.

Shear Testing Programme 2(STEP2) is a project for testing precision of weak lensing measurement (See Massey et al. 2007).
It uses several distortions and 6 PSF patterns for same source objects,
and tested precision of various shear measurement methods including KSB.
Figure 6 of the paper by Massey et. al (2007) 
shows that the precisions are about 1 percent in most of the methods they tested. 
However, Table 4 of the same paper shows that artificial corrections are needed to obtain 
such a precision.
It is natural to imagine that such artificial corrections will change by every real data and every shot of observation, because the artificial corrections corrects the systematic error which are due to size, MAG and SN of objects, size of PSF and so on.
We need some sort of correction in each data, but it will be hard to have universal corrections independent of various observational conditions and various PSFs.
For reducing such hardness, it is thus desirable to develop a new shear measurement methods which need fewer or/and less artificial corrections.

One of the reason to need such corrections is the use of insufficient expansion of the 
weight function as demonstrated in section 4. This lead us to adopt the elliptical 
weight function to measure the galaxy shape with the distortion given by the lensing distortion.  
This will give us a relatively simple relation between the observed, intrinsic shear and lensing distortion. 
In addition, because using same ellipticity of image for weight function increases SN of measurement, 
we can reduce random count noise.

In this section we show the detail of our approach to estimate a lensing shear by using a elliptical weight function.
\subsection{Distortion with weight function}
E-HOLICs method is defined that the weight functions to measure the shape must have no ellipticity in zero planes. 
This directly gives us the following simple expression for spin-2 E-HOLICs without any approximations
according to the similar calculations leading to eq.(\ref{eq:Weightdelta}).
\begin{eqnarray}
\label{eq:NWEHOLICs}
\frac{Z^2_2(I^L,\bdelta)}{Z^2_0(I^L,\bdelta)}\equiv \cH^2_2(I^L,Z^2_0,\bdelta)=\bdelta.
\end{eqnarray}
Moreover, the relations between complex distortion and higher order E-HOLICs can be obtained similarly as eq.(\ref{eq:distHighH}). 
\begin{eqnarray}
\label{eq:distHighHW}
\cH^2_2(I,Z^2_0,\bdelta)&=&\bdelta\\
\cH^4_2(I,Z^4_0,\bdelta)&=&\frac{3         }{2+ \delta^2}\bdelta  \\
\cH^4_4(I,Z^4_0,\bdelta)&=&\frac{3         }{2+ \delta^2}\bdelta^2\\
\cH^6_2(I,Z^6_0,\bdelta)&=&\frac{4+\delta^2}{2+3\delta^2}\bdelta  \\
\cH^6_4(I,Z^6_0,\bdelta)&=&\frac{5         }{2+3\delta^2}\bdelta^2\\
\cH^6_6(I,Z^6_0,\bdelta)&=&\frac{5         }{2+3\delta^2}\bdelta^3
\end{eqnarray}
\subsection{One dimensional form}
In next section, we demonstrate how to correct isotropic PSF.
However because our weight function has ellipticity,
PSF corrections are more complicated than that of KSB methods.
Therefore it is essential to reduce the equations to one dimensional form 
in order to make the method tractable in real observations.
We will explain what is meant by one-dimension in the below.

First, we divide complex distortion and complex moments into absolute value and phase angle.
Let $\phi_\delta$ be a phase angle of $\bdelta$, so 
\begin{eqnarray}
\bdelta\equiv\delta e^{i\phi_\delta}.
\end{eqnarray}
And $\phi^N_M$ be a phase angle of $\cH^N_M$, so
\begin{eqnarray}
\cH^N_M(I^L,Z^2_0,\bdelta)\equiv H^N_M(I^L,Z^2_0,\bdelta)e^{i\phi^N_M},
\end{eqnarray}
where $H^N_M=|\cH^N_M|$.
Owing to eq.(\ref{eq:NWEHOLICs}), we can obtain
\begin{eqnarray}
H^2_2\lr{I^L,Z^2_0,\bdelta}=\delta,\\
\phi^2_2=\phi_\delta.
\end{eqnarray}

\begin{figure*}
\epsscale{0.3}
\plotone{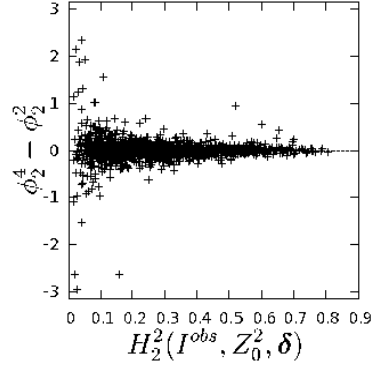}
\caption{
\label{fig:STEP2_ANG42}
Plots of a ellipticity versus a difference of phase angles between $\cH^2_2$ and $\cH^4_2$ of STEP2 objects.
The horizontal axis means an absolute value of ellipticity, and the vertical axis means a phase angle (rad).
Objects that of the difference is close to 0 have high correlation in phase angles.
} 
\end{figure*}

\begin{figure*}
\epsscale{0.3}
\plotone{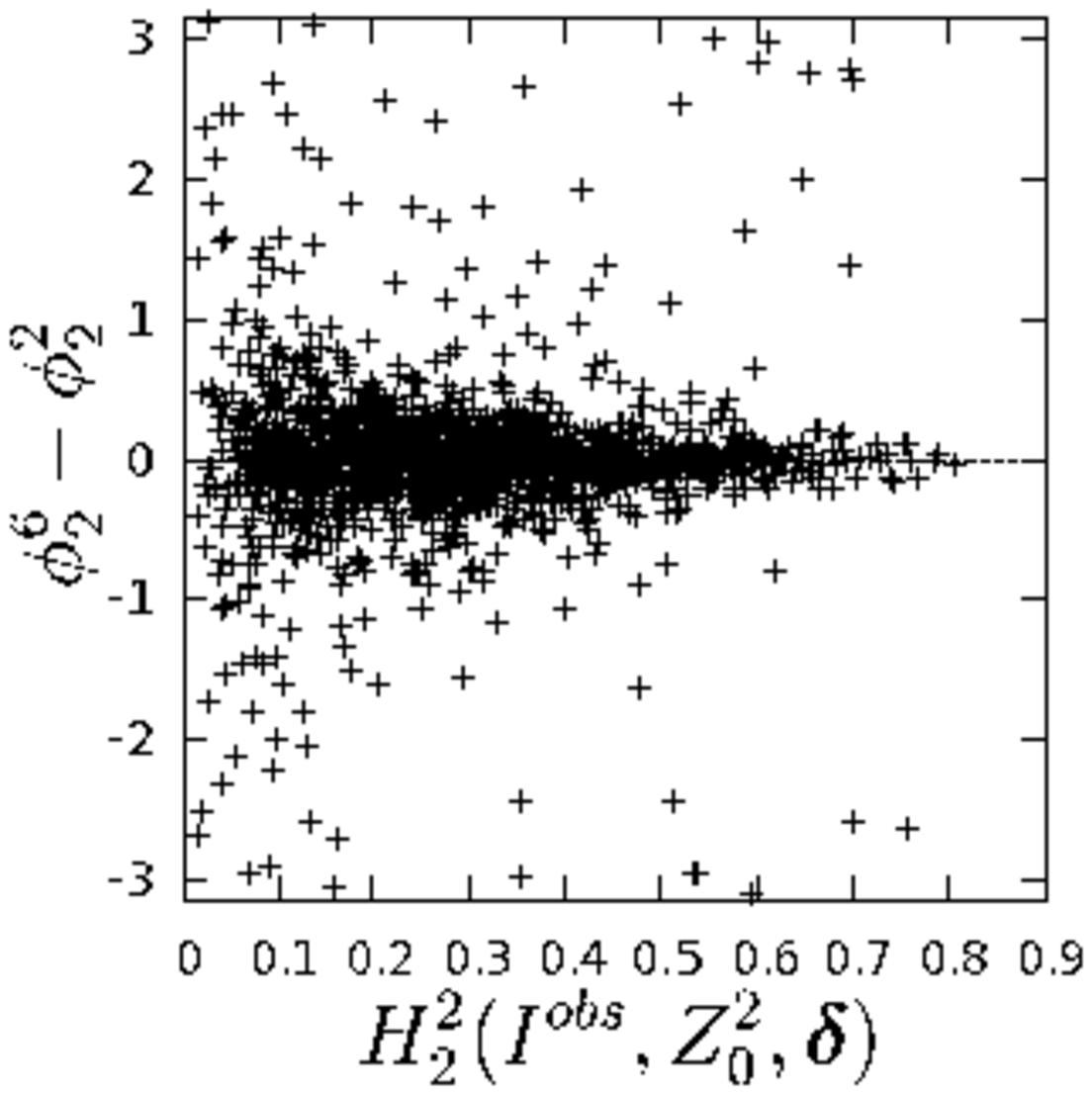}
\caption{
\label{fig:STEP2_ANG62}
Same plots as Figure \ref{fig:STEP2_ANG42} except the difference.
The difference of phase angle between $\cH^6_2$ and $\cH^2_2$ is shown in this figure.
} 
\end{figure*}

\begin{figure*}
\epsscale{0.3}
\plotone{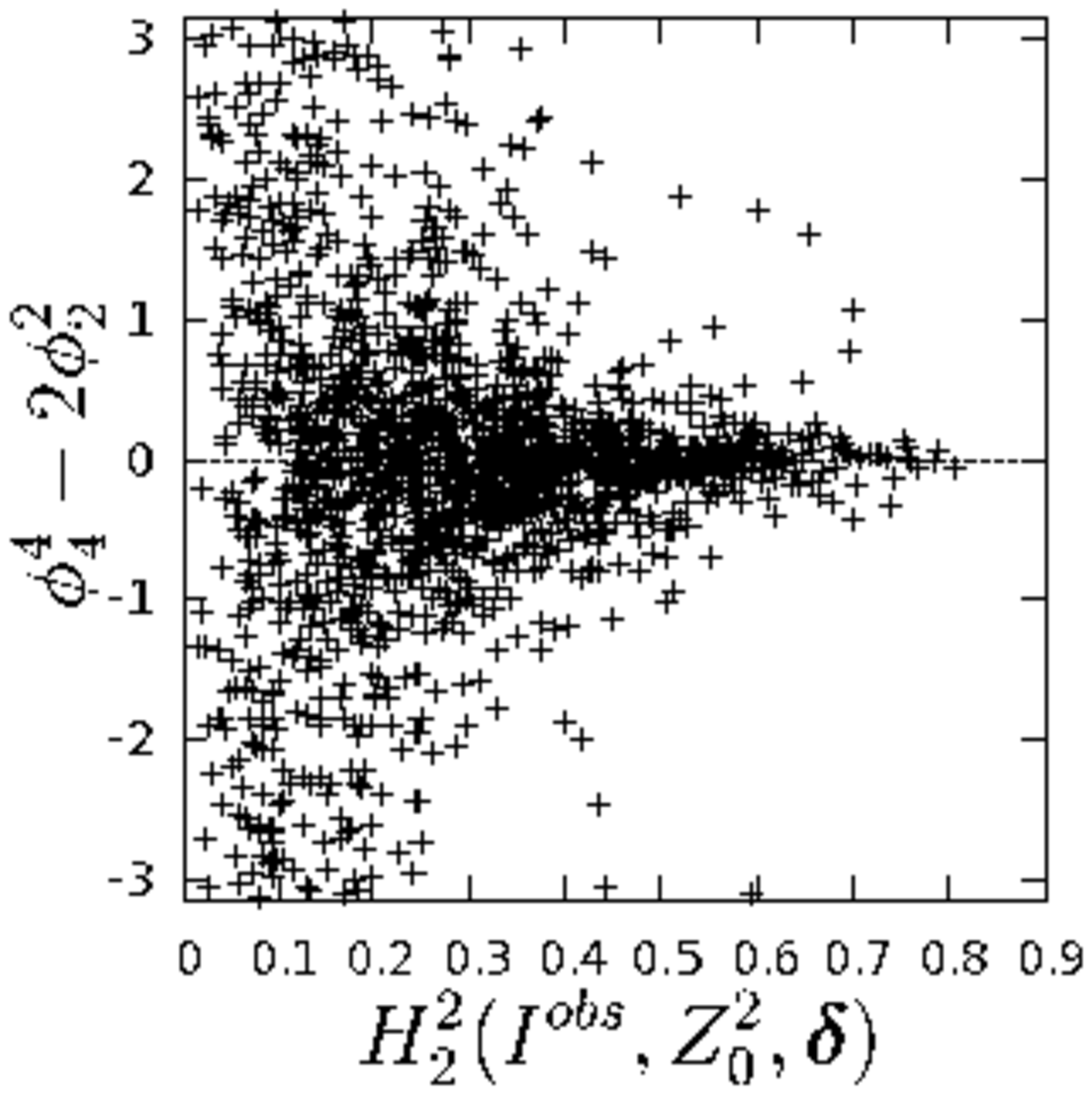}
\caption{
\label{fig:STEP2_ANG44}
Same plots as Figure \ref{fig:STEP2_ANG42} except the difference.
The difference of phase angle between $\cH^4_4$ and twice $\cH^2_2$ is shown in this figure.
} 
\end{figure*}

Next, we calculate the correlations between phase angle of complex distortion and complex moments 
using STEP2(we describe STEP2 simulation images in following section). 
The figures \ref{fig:STEP2_ANG42},  \ref{fig:STEP2_ANG62} show the difference of the phase angle 
between higher-order(N=4, 6) sipn-2 E-HOLICs and complex distortion as a function of ellipticity of images. These figures clearly show a high correlation $(\phi^N_2 \simeq \phi_\delta)$ between them independent of ellipticity of images. 
On the other hand, figure  \ref{fig:STEP2_ANG44} shows the difference of phase between 
spin-4 HOLICs with N=4 and the complex distortion as a function of ellipticity of images. 
This shows also a high correlation $(2\phi^4_4\simeq \phi_\delta)$ at least for images with high ellipticities. 
Fortunately, $\cH^4_4$ appears always as a spin-2 combination with other complex distortion 
(for example $\bdelta^*\cH^4_4$), so terms with $\bdelta^*\cH^4_4$ are always 2nd order or more higher order.
Therefore, although the phase of $\cH^4_4$ has a low correlation with that 
of the lensing distortion if object have low ellipticities, it can safely be neglected 
in our analysis as far as we are interested only in the lowest order calculation. 
Similarly terms containing other higher-order moments with higher spin are  always smaller than 
2nd order. Thus we employ the following approximation for the phase angles of various E-HOLICs. 
\begin{eqnarray}
\phi^N_M\approx\frac{M}{2}\phi^2_2=\frac{M}{2}\phi_\delta.
\end{eqnarray}
This approximation allows us to reduce complex equations for the lensing distortion into real equations of its absolute value. 

\subsection{Isotropic PSF correction}
E-HOLICs method uses similar idea of KSB for PSF correction.
In this paper we present only isotropic PSF correction,
and we will present the anisotropic PSF correction by forthcoming paper.

E-HOLICs method requires Gaussian weight function in the  PSF correction, so
\begin{eqnarray}
W\lr{\frac{\theta^2_0-\Real{\bdelta^*\theta^2_2}}{\sigma^2}}\equiv e^{-\lr{\theta^2_0-\Real{\bdelta^*\theta^2_2}}/\sigma^2}.
\end{eqnarray}
First, we suppose that an isotropic PSF $\tilde P^{iso}(\tilde \beta)$ in the zero plane can be divided into isotropic part and anisotropic part in the image plane as follows.
\begin{eqnarray}
\tilde P^{iso}(\tilde \beta)\equiv\tilde P(\theta)\equiv\int d^2\phi P^{iso}(\phi)\tilde \bq(\theta-\phi).
\end{eqnarray}
where $\tilde P(\theta)$ has the ellipticity which has same value of complex distortion,
and $P^{iso}(\theta)$ is the isotropic part of real PSF.

Let $\tilde I^{iso}$ be the brightness distribution which is smeared by $\tilde P^{iso}(\tilde \beta)$,
then the complex moments of $\tilde I^{iso}$ can be expressed by the complex moments of $I^{iso}$ and $\tilde q$ as
\begin{eqnarray}
\label{eq:isoEH}
\int &d^2\tilde\beta& f(\tilde\beta)\tilde I^{iso}(\tilde\beta) \equiv
\int d^2\tilde\beta f(\tilde\beta)\int d^2\tilde\psi \tilde P^{iso}(\tilde \beta-\tilde \psi)I^Z(\tilde \psi)\\
&=&(1-\kappa)^2(1-|g|^2)\int d^2\theta f(\theta-g\theta^*)\int d\varphi I^L(\varphi)\tilde P(\theta-\varphi)\\
&=&(1-\kappa)^2(1-|g|^2)\int d^2\theta f(\theta-g\theta^*)\int d\varphi I^L(\varphi)\int d^2\phi P^{iso}(\theta-\varphi-\phi)\tilde \bq(\phi)\\
&=&(1-\kappa)^2(1-|g|^2)\int d^2\psi \int d^2\phi f((1-\kappa)(\psi-g\psi^*+\phi-g\phi^*))I^{iso}(\psi)\tilde \bq(\phi).
\end{eqnarray}
where $f((1-\kappa)\lr{\psi-g\psi^*+\phi-g\phi^*})$ is
\begin{eqnarray}
&&f((1-\kappa)\lr{\psi-g\psi^*+\phi-g\phi^*})\\
&&\equiv (1-\kappa)^N(\psi-g\psi^*+\phi-g\phi^*)^N_MW\lr{\frac{(\psi+\phi)^2_0-\Real{\bdelta^*(\psi+\phi)^2_2}}{\sigma^2}},
\end{eqnarray}
and W is expanded by using the property of Gaussian as
\begin{eqnarray}
&&\hspace{-2cm}W\lr{\frac{(\psi+\phi)^2_0-\Real{\bdelta^*(\psi+\phi)^2_2}}{\sigma^2}}\\
																		&=&W\lr{\frac{\psi^2_0-\Real{\bdelta^*\psi^2_2}}{\sigma^2}}W\lr{\frac{\phi^2_0}{\sigma^2}}e^{-\frac{1}{\sigma^2}\lr{\psi\phi^*+\psi^*\phi-\bdelta^*\psi\phi-\bdelta\psi^*\phi^*-\frac{1}{2}\bdelta^*\phi^2_2-\frac{1}{2}\bdelta{\phi^2_2}^*}}\\
																		&\approx&W\lr{\frac{\psi^2_0-\Real{\bdelta^*\psi^2_2}}{\sigma^2}}W\lr{\frac{\phi^2_0}{\sigma^2}}\times\\
            															&&\biggl(1-\frac{1}{\sigma^2}\lr{\psi\phi^*+\psi^*\phi-\bdelta^*\psi\phi-\bdelta\psi^*\phi^*-\frac{1}{2}\bdelta^*\phi^2_2-\frac{1}{2}\bdelta{\phi^2_2}^*}\\
            															&&+\frac{1}{2\sigma^4}\lr{\psi\phi^*+\psi^*\phi-\bdelta^*\psi\phi-\bdelta\psi^*\phi^*-\frac{1}{2}\bdelta^*\phi^2_2-\frac{1}{2}\bdelta{\phi^2_2}^*}^2\biggr).
\end{eqnarray}
Here, the complex moments of $\tilde q$ is defined as
\begin{eqnarray}
\tilde \bq^N_M\equiv\frac{\int d^2\theta \theta^N_MW\lr{\frac{\theta^2_0}{\sigma^2}}\tilde q(\theta)}{\int d^2\theta W\lr{\frac{\theta^2_0}{\sigma^2}}\tilde q(\theta)},
\end{eqnarray}
where $\tilde q^N_M$ satisfies $\tilde q^0_0\equiv 1$ and $\tilde q^N_0\equiv 0$($N\neq0$),
and we assume that $\tilde \bq^0_0$ and $\tilde \bq^2_2$ are dominant and we ignore other moments of $\tilde q$. 
Moreover, because the phase angle of $e^{i\phi_q}$(defined as $\tilde \bq^2_2\equiv\tilde q^2_2 e^{i\phi_q}$) is same as $\phi_\delta$, 
eq.(\ref{eq:isoEH}) reduces to one dimensional form and we obtain
\begin{eqnarray}
0&=&Z^2_2(\tilde I^{iso},0)\approx(1-\kappa)^4(1-|g|^4)\Biggl[Z^2_2-\bdelta Z^2_0\\
&+&\tilde \bq^2_2\lr{Z^0_0-\frac{1}{2\sigma^2}\lr{(4+3\delta^2)Z^2_0-7\bdelta^*Z^2_2}+\frac{1}{2\sigma^4}\lr{(1+2\delta^2)Z^4_0-\bdelta^*(2+\delta^2)Z^4_2-\bdelta Z^{4*}_2+\bdelta^{*2} Z^4_4}}\\
&+&\tilde \bq^{2*}_2\lr{-\frac{1}{2\sigma^2}(3\bdelta^2Z^2_0-3\bdelta Z^2_2)+\frac{1}{2\sigma^4}(3\bdelta^2Z^4_0-3\bdelta Z^4_2+\bdelta^3Z^{4*}_2+Z^4_4)} \Biggr](I^{iso},\bdelta)\\
&\approx&(1-\kappa)^4(1-|g|^4)Z^2_0(I^{iso},\bdelta)e^{i\phi_\delta}\Biggl[H^2_2-\delta+\tilde q^2_2\Bigl(H^0_0-\frac{1}{\sigma^2}\lr{(2+3\delta^2)H^2_0-5\delta H^2_2}\\
&&+\frac{1}{2\sigma^4}\lr{(1+5\delta^2)H^4_0-2\delta(3+\delta^2)H^4_2+(1+\delta^2)H^4_4}\Bigr)\Biggr](I^{iso},Z^2_0,\bdelta)\\
&\equiv&(1-\kappa)^4(1-|g|^4)Z^2_0(I^{iso},\bdelta)e^{i\phi_\delta}\Bigl(H^2_2(I^{iso},Z^2_0,\bdelta)-\delta+\tilde q^2_2P^E(I^{iso},\bdelta)\Bigr),
\end{eqnarray}
where $P^E$ is a combination of absolute values of HOLICs and $\bdelta$, thus $P^E$ is real. 
Therefore, from the above equation, we have the equation for $\delta$.
\begin{eqnarray}
\delta \approx H^2_2(I^{iso},Z^2_0,\bdelta)+\tilde q^2_2P^E(I^{iso},\bdelta).
\end{eqnarray}
This equation can be apply to star objects to have an expression for  $\tilde q^2_2$. 
\begin{eqnarray}
\tilde q^2_2\approx\frac{\delta-H^2_2(I^{*iso},Z^2_0,\bdelta)}{P^E(I^{*iso},\bdelta)},
\end{eqnarray}
where $I^{*iso}$ is circle but $H^2_2$ does not vanish because our weight function has ellipticity.
Finally, the lensing distortion can be evaluated by solving the following equation
\begin{equation}
\label{eq:EHdist}
\delta\approx \delta^E\equiv H^2_2(I^{iso},Z^2_0,\bdelta)+\frac{P^E(I^{iso},\bdelta)}{P^E(I^{*iso},\bdelta)}\lr{\delta-H^2_2(I^{*iso},Z^2_0,\bdelta)}
\end{equation}
The complex distortion then is obtained as 
\begin{eqnarray}
\bdelta\approx\bdelta^E&\equiv&\lr{H^2_2(I^{iso},Z^2_0,\bdelta)+\frac{P^E(I^{iso},\bdelta)}{P^E(I^{*iso},\bdelta)}\lr{\delta-H^2_2(I^{*iso},Z^2_0,\bdelta)}}e^{i\phi^2_2}\\
&\approx&\lr{\cH^2_2(I^{iso},Z^2_0,\bdelta)+\frac{P^E(I^{iso},\bdelta)}{P^E(I^{*iso},\bdelta)}\lr{\bdelta-\cH^2_2(I^{*iso},Z^2_0,\bdelta)}}
\end{eqnarray}
\subsection{Ellipticity of weight function}
Eq.(\ref{eq:EHdist}) requires us to use the ellipticity of lensing distortion. 
However an observed object has a different ellipticity from the lensed ellipticity due to PSF.
Therefore we can not obtain lensing distortion directly by observation, and thus 
we use temporally the observed ellipticity as a natural first step. 
Then there are two ways to estimate the lensing distortion.
One is iteration. Namely we use the observed ellipticity for elliptical weight and re-estimating lensed ellipticity, and iterate measurement by using re-estimated ellipticity until the result converges.
Otherwise we could expand the weight function for all moments of  Eq.(\ref{eq:EHdist}) as
\begin{eqnarray}
\hspace{-3cm}  &W &\lr{ \frac{\theta^2_0-\Real{\bdelta^*\theta^2_2}}{\sigma^2}} \\
&\approx& W\lr{\frac{\theta^2_0-\Real{\bdelta^*_W\theta^2_2}}{\sigma^2}}\Biggl(1+\frac{1}{2\sigma^2}\lr{\lr{\bdelta-\bdelta_W}\theta^{2*}_2+\lr{\bdelta^*-\bdelta^*_W}\theta^2_2}\\
&&+\frac{1}{4\sigma^4}\lr{\lr{\bdelta-\bdelta_W}^2\theta^{4*}_4+\lr{\bdelta-\bdelta_W}\lr{\bdelta^*-\bdelta^*_W}\theta^4_0+\lr{\bdelta^*-\bdelta^*_W}^2\theta^4_4}\Biggr),
\end{eqnarray}
where $\bdelta_W$ is an ellipticity we use for weight function.
It seems reasonable to think that coefficients of this expansion are smaller than these of KSB method. 
By using observed ellipticity for $\bdelta_W$,
systematic error due to this expansion will be smaller than that in KSB. 

We have examined both method and found the iteration needs more time than the expansion. 
Thus the above expansion of the weight function may be more useful in real observation 
because we must estimate PSF effects for many star objects. 

Two dimensional fitting set of $\tilde q^2_2$ with elliptical weight functions which have all phase angle and ellipticity or
using only some star objects which surround each distorted object which we try to correct PSF effects will be effective in saving time to calculate.
\section{STEP2 simulation test}
In this section, we perform tests of our methods and compare our results with KSB method by using STEP2 simulation data.
Because we presented only isotropic PSF correction, 
we use data set of psf-F which have no anisotropic part. 
There are 64 data set of STEP2 simulation data which are distorted by different distortion respectively and have a rotated image. 
By using these rotated data, an effect of intrinsic ellipticity can be canceled
 (but a systematic error of intrinsic ellipticity can not be canceled completely).
In our evaluation of shear we have ignored denominator of eq.(\ref{eq:gICL}) when averaging for neglecting systematic error. 
In real analysis, we should use iteration to consider higher order in the equation.
A perl scripts (Umetsu Keiichi private communication) and IMCAT were used for detecting objects.
In averaging with the normal and the rotated data, 
the number of matched objects in No.36 field in STEP2 is about 1/10 of the number of matched objects in other fields(may be due to some strangeness in detection), thus 
we neglected the result of the field.
\subsection{Star selection}
We selected star objects by the condition 1.9$<r_h<$2.2(pix), 20$<$MAG$<$23.
Generally, two dimensional fitting is used for PSF parameters in real weak lensing analysis,
because PSF effect slightly changes in a data field. 
However, constant PSF is used in STEP2 simulation data,
so we can correct PSF effects by using only one star object.
Here we use one of good star which has SN$\approx$500 for PSF correction.
In order to see the detailed behavior of the estimation depending on the choice of the objects, 
we classify the objects into the following classes. 

\subsection{good objects}
First we define a class which contains only good objects. Good objects is defined to 
be objects which are large ($r_h>$3.3(pix)) and have high SN (SN$>$20) and 
small intrinsic ellipticities($|\delta_{intrinsic}|<$0.35). 
We do not use any limit on MAG(e.g. 20$<$MAG$<$26). 
These conditions reduces objects used for the shear measurement where KSB method have a number density 1.50(/min$^2$) and E-HOLICs method have a number density 1.68(/min$^2$).
Large images have fewer effect from PSF and high SN images have fewer effect from random 
count noise. The effect of expansion of weight function will not cause large error in shape measurement for images with small intrinsic images. Thus the estimated shear using only good object is expected to be 
close to the input value.   

\begin{figure*}
\epsscale{1.0}
\plotone{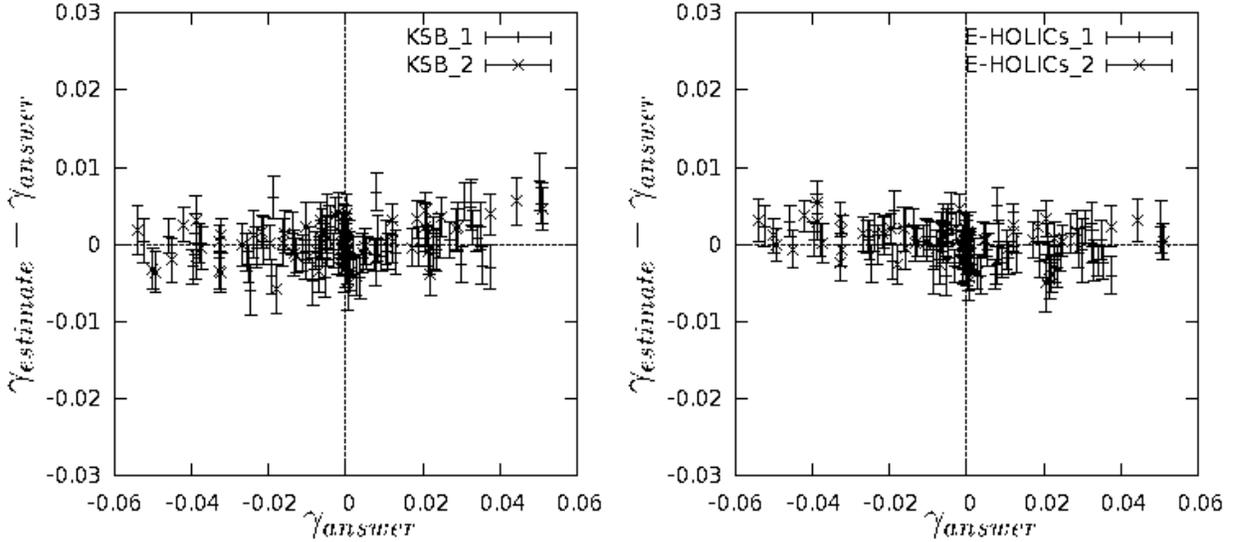}
\caption{
\label{fig:STEP2_good}
Plots of results of test of 63 fields of STEP2,
The left side shows a result of KSB method and the right side shows a result of E-HOLICs method.
The horizontal axis means value of inputted shear(answer), 
and the vertical axis means a value of difference between estimated shear and inputted shear.
The subscripts 1 and 2 mean components of directions.
KSB method has a number density 1.50(/min$^2$) and obtained fitting parameter 0.0386$\gamma$+0.000232.
E-HOLICs method has a number density 1.68(/min$^2$) and obtained fitting parameter -0.0242$\gamma$+0.000083.
} 
\end{figure*}

Figure \ref{fig:STEP2_good} shows results of shear estimation by KSB method and E-HOLICs method 
using only good objects. In these figures the horizontal axis means inputted(answer) shear and the vertical axis means a difference between estimated shear and inputted shear.

The result is described by two parameters m and c defined as $\gamma_{e}=m \gamma_{i} +c $ 
where $\gamma_{e}$ is the estimated shear and $\gamma_{i}$ is the input shear. 
KSB gives m=0.03868 and c=0.000232, and E-HOLICs gives  m=$-$0.0242 and c=0.000083. 
Both give satisfactory result even if we used only a few number of images.
Comparing with both methods, we can see KSB method gives slight over estimation for objects 
with large ellipticities. 

\subsection{intrinsic ellipticity, SN and size dependence}
Next we would like to see the effect of intrinsic ellipticity, SN and size of objects separately in the shear estimation.

\begin{figure*}
\epsscale{1.0}
\plotone{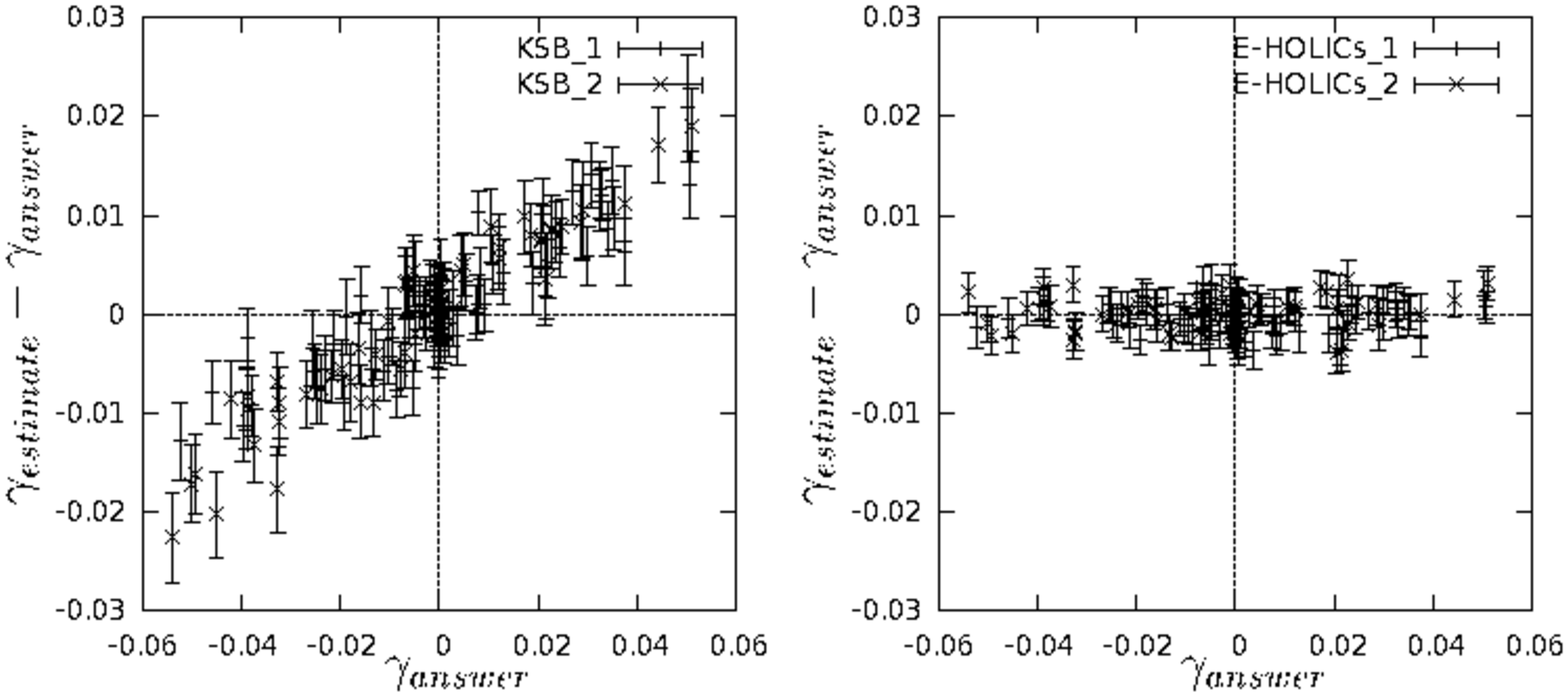}
\caption{
\label{fig:STEP2_ALLe}
Same plots as Figure \ref{fig:STEP2_good} except the limit of ellipticity for using objects,
and the limit of this figure is $|\delta|<$1.
KSB method has a number density 2.48(/min$^2$) and obtained fitting parameter 0.3108$\gamma$+0.000737.
E-HOLICs method has a number density 3.41(/min$^2$) and obtained fitting parameter -0.0044$\gamma$+0.000104.
} 
\end{figure*}

First we relax the condition for intrinsic ellipticity keeping other conditions. 
We include object with  $|\delta_{intrinsic}|<$1.  
This increases the number of objects up to 2.48(/min$^2$) for KSB and 3.41(/min$^2$) for E-HOLICs.
Figure \ref{fig:STEP2_ALLe} shows the results where  
KSB gives m=0.3108 and c=0.000737 and E-HOLICs gives m=0.0044 and c=$-$0.000104. 
We can see that KSB method overestimates the input value strongly, 
The estimation by E-HOLICs method does improve, but the degree of improvement in this test 
is not explained by the increase of number density. It seems the this 
increase is explained by statistical variance. 
This is an evidence of that the expansion of weight function in KSB method is insufficient.
Moreover, the result by KSB method has larger error bar compared to E-HOLICs method. This is 
because systematic error due to ellipticity does not cancel completely.

\begin{figure*}
\epsscale{1.0}
\plotone{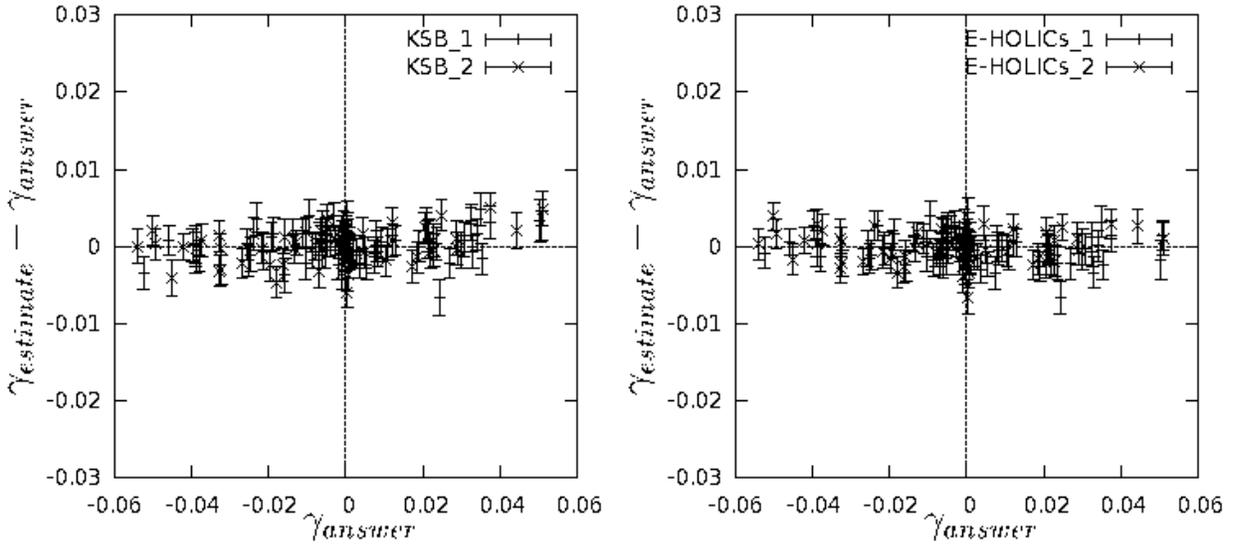}
\caption{
\label{fig:STEP2_ALLrh}
Same plots as Figure \ref{fig:STEP2_good} except the limit of $r_h$ for using objects,
and the limit of this figure is 2.2$<r_h$.
KSB method has a number density 3.85(/min$^2$) and obtained fitting parameter  0.0277$\gamma$+0.000231.
E-HOLICs method has a number density 4.05(/min$^2$) and obtained fitting parameter -0.0075$\gamma$-0.000233.
} 
\end{figure*}

To see the effect of size dependence in the shear estimation, 
we relax the condition of the size 
keeping other conditions fixed. We include objects with 2.2$<r_h$ on the top of good objects, and 
this increases number density of the objects up to 3.85(/min$^2$) 
for KSB and 4.05(/min$^2$) for E-HOLICs.  
Figure \ref{fig:STEP2_ALLrh} shows results where   
KSB gives m=0.0277 and c=0.000231 and E-HOLICs gives $m=-0.0075$ and $c=-0.000233$.
We can see there is no strong effects from $r_h$ dependence in both method.
The reason for huge improvement may also be due to statistical variance.  

Finally we examine the effect of SN in the shear measurement. We therefore relax the condition for 
SN keeping other conditions fixed. We include objects with SN$>$2 on the top of good objects, which 
increases the number density up to 4.99(/min$^2$) for KSB and 5.38(/min$^2$) for E-HOLICs.

\begin{figure*}
\epsscale{1.0}
\plotone{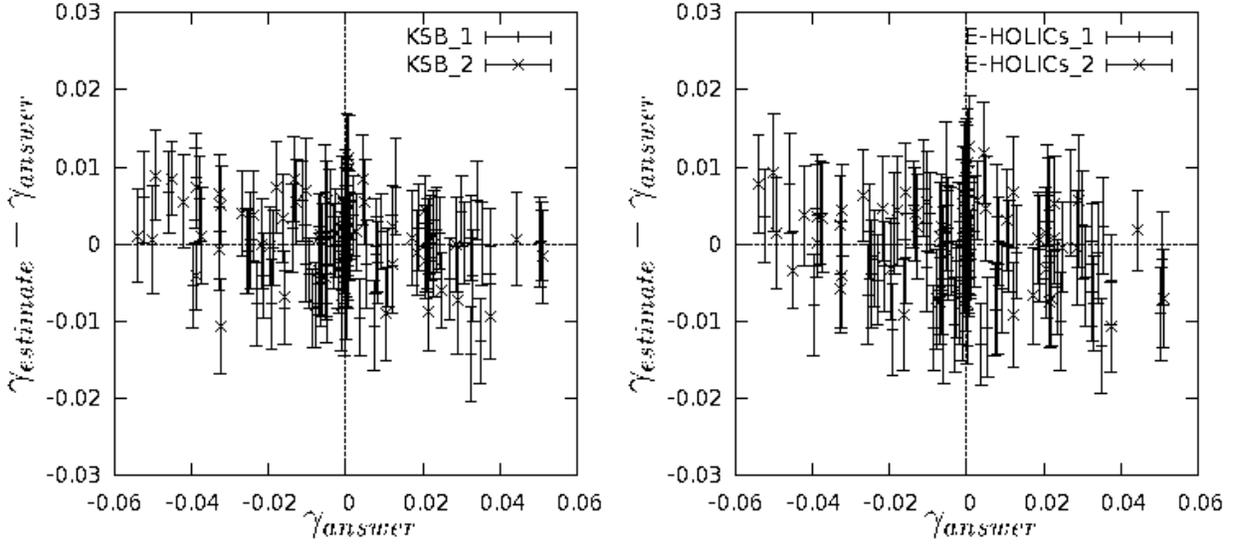}
\caption{
\label{fig:STEP2_SN2}
Same plots as Figure \ref{fig:STEP2_good} except the limit of SN for using objects,
and the limit of this figure is 2$<$SN.
KSB method has a number density 4.99(/min$^2$) and obtained fitting parameter -0.0539$\gamma$-0.000105.
E-HOLICs method has a number density 5.38(/min$^2$) and obtained fitting parameter -0.0604$\gamma$-0.000553.
} 
\end{figure*}

Figure \ref{fig:STEP2_SN2} shows results where 
KSB gives m=$-$0.0539 and c=0.000105 and E-HOLICs gives m=$-$0.0604 and c=$-$0.000553. 
We can see both methods underestimate the input value slightly and gain larger errors. 
This may be explained by the fact that random count noise reduce the observed ellipticities by adding random ellipticities.
Because the degree of reduction by KSB method is larger than that of E-HOLICs method, one can conclude that KSB method is more affected from random count noise than E-HOLICs methods.

We also found that the use of stars with low SN underestimates the input value, 
because random count noise disturb a precise measurement of an effect of PSF.
Unfortunately, using only high SN stars is incompatible with small scale PSF correction.
Because this problem is related with anisotropic PSF correction,
we will address this problem in the forthcoming paper.

\subsection{ALL object}
\begin{figure*}
\epsscale{1.0}
\plotone{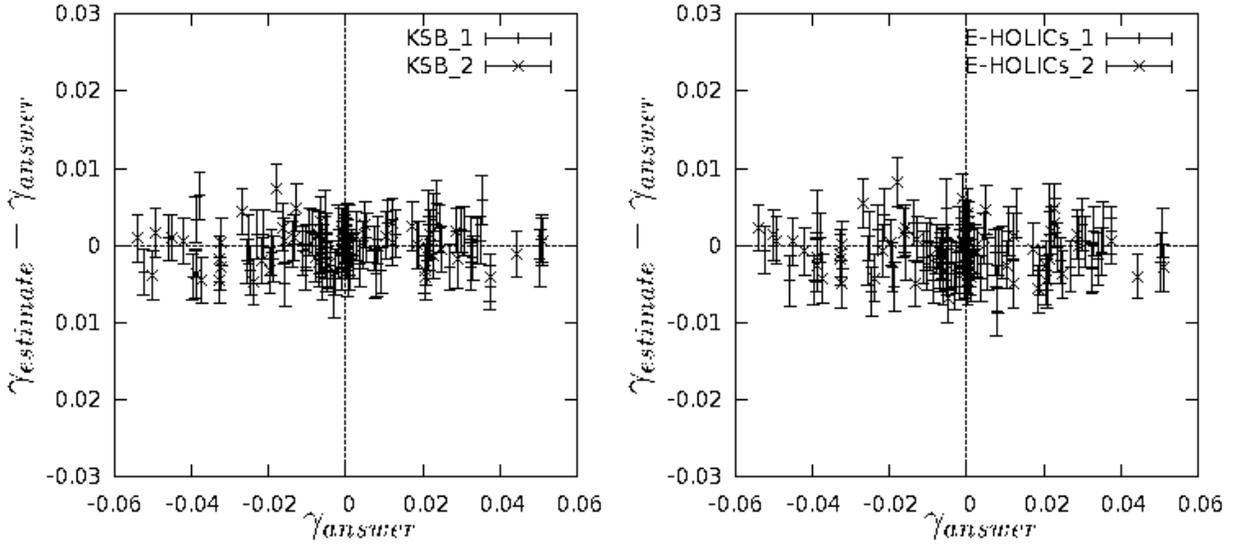}
\caption{
\label{fig:STEP2_ALL}
Same plots as Figure \ref{fig:STEP2_good} except the limits for using objects,
and the limits of this figure is $|\delta|<$1, 2.2$<r_h$ and 4$<$SN.
KSB method has a number density 23.70(/min$^2$) and obtained fitting parameter 0.0099$\gamma$-0.000261.
E-HOLICs method has a number density 24.42(/min$^2$) and obtained fitting parameter -0.0014$\gamma$-0.000863.
} 
\end{figure*}

Finally, figure \ref{fig:STEP2_ALL} shows the test using all object without any conditions mentioned above. Namely we use objects with  2.2$<r_h$、$|\delta|<$1, 4$<$SN. 
This will increase the number density available for the measurement up to 23.70(/min$^2$) for KSB 
and 24.42(/min$^2$) for E-HOLICs. 
The result shows that m=0.0099 and c=$-$0.000261 in KSB method and m=$-$0.0014 and c=$-$0.000863 in E-HOLICs method. 
We can see both methods give good estimation for input shear in this case. 
However, it should be mentioned that the accurate estimation obtained by KSB method 
is the result of fortunate cancelation between overestimate for images with large ellipticities and underestimate for images with low SN. 
On the other hand, because E-HOLICs method improves only  overestimation due to the expansion of 
weight function, E-HOLICs method has a tendency to slightly underestimate the input shear.
\section{Summary}
We have developed a new method to estimate gravitational shear, called "E-HOLICs method"  based on KSB method by introducing the elliptical weight to define multipole moments of galaxy light distribution. 
In E-HOLICs method we use the lensed ellipticity for the ellipticity of weight function, and thus 
it avoids a systematic error coming from an expansion of weight function which is done in previous 
approaches for shape measurement including KSB method. 
Furthermore the elliptical weight measures the galactic shape more effectively than with circular weight function, and it is expected to reduces random count noise. 
 
We have also developed the isotropic PSF correction for E-HOLICs method and were able to show that 
the equation which governs the correction reduces to real polynomial for the magnitude of the lensing distortion. 
This is possible because there is a high correlations between the phase angle of various 
order of E-HOLICs with same spin.  For the application to real observation we need to correct anisotropic part of the effect PSF which will be discussed in the forthcoming paper. 

We have also introduced a new concept of "zero plane" in weak lensing analysis which is an imaginal source plane where shape of all sources is perfect circle, and the intrinsic shear is regarded as the result of an imaginal lensing distortion. This makes the relation between the observed shear, the intrinsic shear and lensing distortion simple, and  higher-order analysis easy. The idea of zero plane may be used in any shear estimation scheme. 

We have performed various tests to show the improvements of shear measurement by E-HOLICs 
method using STEP2 simulation data, and compared the results with those obtained by using KSB method. 
We have examine effects of ellipticity, size and SN of images separately in both method, 
and found that E-HOLICs method can reduce the errors coming from these effects. 
In particular, we have confirmed that KSB method overestimates the shear if images with large ellipticities are used in the analysis, and E-HOLICs method does not have such tendency. 

Although E-HOLICs method has  a potentiality to accurately measure the shear, it requires 
more time for the estimation of the shear than KSB method in general, 
because it requires iteration to measure the multipole moments of background galaxies. 
Therefore it will be necessary to develop more effective method for the shape measurement 
when we apply E-HOLICs method to planned large scale cosmic shear observations.

\acknowledgments

We thank 
K. Umetsu for a useful discussion and providing his scripts,
T. Okamura for helping us in calculation,
T. Hamana and M. Oguri for a useful discussion. This work is supported in part by the GCOE Program "Weaving Science Web beyond Particle-matter Hierarchy" at Tohoku University and by a Grant-in-Aid for Scientific Research from JSPS(Nos. 18072001, 20540245 for TF) as well as by Core-to-Core Program "International Research Network for Dark Energy".
\begin{eqnarray}
\end{eqnarray}

\end{document}